# Analysis of Evidence using Formal Event Reconstruction

Joshua James[1], Pavel Gladyshev, Mohd Taufik Abdullah, Yuandong Zhu
Joshua.James@UCD.ie, Pavel.Gladyshev@UCD.ie, Yuandong.Zhu@UCD.ie

Centre for Cybercrime Investigation
University College Dublin
Belfield, Dublin 4, Ireland

## Abstract
This paper expands upon the finite state machine approach for the formal analysis of digital evidence. The proposed method may be used to support the feasibility of a given statement by testing it against a relevant system model. To achieve this, a novel method for modeling the system and evidential statements is given. The method is then examined in a case study example.

## Introduction
A sound forensic analysis is expected to rely on a credible scientific theory that explains why and how expert conclusions follow from the available evidence (Gladyshev & Patel, Finite State Machine Approach to Digital Event Reconstruction, 2004). While advancements have been made in the formalization of the digital investigation process, analysis still remains largely ad-hoc. This paper adds to a body of work that attempts to formalize the digital investigation process, and specifically works to extend the finite state machine (FSM) theory (Carrier & Spafford, Categories of digital investigation analysis techniques based on the computer history model, 2006)(Gladyshev, Finite State Machine Analysis of a Blackmail Investigation, 2005) (Gladyshev & Patel, Finite State Machine Approach to Digital Event Reconstruction, 2004).

As stated in (Gladyshev, Finite State Machine Analysis of a Blackmail Investigation, 2005), "many digital systems, such as digital circuits, computer programs, and communication protocols can be described mathematically as finite state machines. A finite state machine can be viewed as a graph whose nodes represent possible system states, and whose arrows represent possible transitions from state to state". Likewise, Carrier claims, "modern computers are FSMs with a large number of states and complex transition functions"(Carrier, A Hypothesis-Based Approach to Digital Forensic Investigations, 2006). By utilizing this fact, finite state machine models that represent the computations of a system may be formally defined. It is then possible to test scenarios in terms of the model to see if given situations are computationally possible.

## Formal Methods of Investigation
Other methods for the formalization of the digital investigation process have previously been proposed, such as Semantic Integrity Checking (Stallard & Levitt, 2003), Temporal Logic of Security Actions (Rekhis, 2008), and previous Finite State Machine modeling approaches (Gladyshev, Finite State Machine Analysis of a Blackmail Investigation, 2005), among others. Semantic Integrity Checking (Stallard & Levitt, 2003) involves the analysis of redundant data objects that must exist in a system. From found inconsistencies in the redundant data it is possible to hypothesize attack scenarios. The drawback to this technique is that if there is a lack of data with which to corroborate events (no IDS logging, disabled firewall logging, etc), such as in standard home computers, it is entirely possible that an attacker could forge or simply remove required redundant traces. This dependency on the existence of redundant data is useful in environments with strict logging and security policies, but may have issues reliably scaling down to the level of the average home computer. Temporal Logic of Security Actions (S-TLA) is a Logic-Based Language for Digital Investigations (Rekhis, 2008) which represents the system using state-based logic. By combining pre-defined generic scenario fragments with the created system model, undesirable states may be found from which evidence may be derived. From the combination of recurring scenario fragments and found evidence, formulation of possible event scenarios can occur. A weakness of this method comes from the definition of scenario fragments. These must be pre-defined, and are unable to be automatically generated given a suspect system. This means that if a fragment is not defined for a certain action, or set of actions, then undesirable states correlating to this fragment would not be identified. Without the definition of fragments for each possible action, at most, a partial view of the total action and evidence would be considered. The Finite State Machine approach proposed by (Gladyshev & Patel, Finite State Machine Approach to Digital Event Reconstruction, 2004) models the system as an FSM whose transitions may be back-traced from the state the system was found. Witness observations are used to restrict the possible transitions of the system. From this restricted back-tracing, possible incident scenarios may be found. Since this method considers each possible transition in the model, exponential growth of the state space is an issue. This greatly limits the methods ability to model complex real-world systems.

[1] Research funded by the Science Foundation Ireland (SFI) under Research Frontiers Programme 2007 grant CMSF575

**Contribution**

This paper expands upon the idea of using formal analysis to test the feasibility of a given witness statement. To do this, a novel approach to formally defining the system is given. An algorithm is proposed to represent the system as a deterministic finite automaton (DFA) that encodes the set of system computations as a set of strings. Witness statements are then formally defined as restrictions on strings accepted by the DFA.

**Organization**

The remainder of this paper is comprised of four sections. In the first section an informal overview of the method is given. The second section explains how to derive a DFA representative of the computations of the system. The third section applies automata intersection using both the system and witness statement models to test validity of the given witness statement. This process is illustrated in a given case study. Finally, considerations of the strengths and weaknesses of the proposed method will be given.

## I. Representing the System

For a given Finite State Machine each transition can be encoded as a triple (state1, input, state2). For example, in figure 1 there are two transitions; A-1->B and B-1->B.

They can be encoded as [A, 1, B] and [B, 1, B], respectively. A sequence of transitions may be defined as a computation. An example of which is the transition 'A-1->B-1->B' that may be expressed as ([A, 1, B][B, 1, B]).

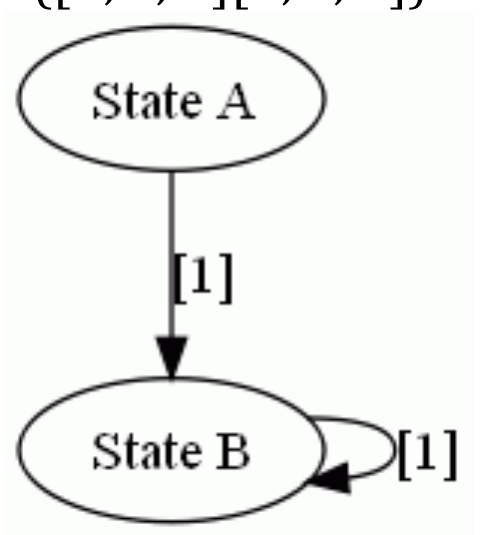

**Figure 1 - Simple Finite State Machine**

Assuming that a computation can start in any state, the entire set of possible computations of the FSM in figure 1 is as follows:

**A = {** [A, 1, B]
[A, 1, B] [B, 1, B]
[A, 1, B] [B, 1, B] [B, 1, B]
[A, 1, B] [B, 1, B] [B, 1, B] [B, 1, B]
...
[B, 1, B]
[B, 1, B] [B, 1, B]
[B, 1, B] [B, 1, B] [B, 1, B]
... **}**

**Witness Statements**

Witness statements can be viewed as restrictions on possible computations of the FSM (Gladyshev & Patel, Finite State Machine Approach to Digital Event Reconstruction, 2004). These restrictions can be expressed as regular expressions (Warren, Regular Expressions, 1999), or patterns over the sequences of transition triples. For example, an observation that the FSM from figure 1 started in 'State A', can be written as:

[A, 1, B]* †

which corresponds to all possible sequences of triples that begin with [A, 1, B]:

**B = {** [A, 1, B]
[A, 1, B] [A, 1, B]
[A, 1, B] [B, 1, B]
[A, 1, B] [A, 1, B] [A, 1, B]
[A, 1, B] [A, 1, B] [B, 1, B]
[A, 1, B] [B, 1, B] [A, 1, B]
[A, 1, B] [B, 1, B] [B, 1, B]
... **}**

Observe that not all of the above sequences correspond to valid computations of the FSM in figure 1. The subset of computations of the FSM from figure 1 that obey the restriction † can be obtained by intersecting

sets **A** and **B**:

**A ∩ B = {** [A, 1, B]
[A, 1, B] [B, 1, B]
[A, 1, B] [B, 1, B] [B, 1, B]
... **}**

Although sets **A** and **B** are infinite, it is possible to construct finite automata that represent them.

## II. Constructing a DFA of the System Model

In this section the system is represented as its corresponding Finite State Machine ($M$). From $M$ a DFA may be derived that accepts the computations of $M$ as a set of strings. Witness statements and observations are also described. A method for dealing with partial observations within a witness statement is given.

### Definition of the System

The given system may be directly mapped as an FSM. Following (Gladyshev & Patel, Finite State Machine Approach to Digital Event Reconstruction, 2004), a finite state machine is defined as a triple $M = (Q, \Sigma, \delta)$, where

- $Q$ is the finite set of all possible states
- $\Sigma$ is the finite set of all possible event
- $\delta : Q \times \Sigma \rightarrow Q$ is a transition function that returns the next state. $\delta$ is a total function.

This definition of an FSM is considered to be deterministic because "... the next state is uniquely determined by a single input event" (Parker).

### Representing the System Model

Given $M$, the proposed algorithm produces a DFA ($M_1$) that represents the set of computations of $M$ encoded as a set of strings. Essentially this can be thought of as a DFA that accepts the computations that $M$ performs (figure 2). Formally $M_1$ is defined as $M_1 = (Q_1, \Sigma_1, \delta_1, g, F)$, where

- $Q_1$ is the finite set of all possible states
- $\Sigma_1$ is the newly created finite set of all possible events
- $\delta_1 : Q_1 \times \Sigma_1 \rightarrow Q_1$ is a transition function that determines the next state of $M_1$
- $g$ is the start state where $g \notin Q$
- $F$ is the set of accepting states

### Constructing a DFA of the System Model

Given $M$:

1. Define $\Sigma_1$ as a finite set of events where each event in $\Sigma_1$ is a triple $\varphi = (q, e, q_2)$. $\Sigma_1 = \{\varphi \mid \varphi = (q, e, q_2)\}$, where
   - $q \in Q$
   - $e \in \Sigma$
   - $q_2 = \delta(q, e)$

2. Define $Q_1$ as containing the set $Q$ as well as a generic start state $g$ where $g$ is not in the set $Q$.
   ***Define:*** $\boldsymbol{Q_1 = Q \cup \{g\}, g \notin Q}$

3. Define a transition function $\delta_1$ where for each transition in the set of $\delta$ there also exists a transition in $\delta_1$ from $q$ to $q_2$ and from $g$ to $q_2$ labeled "$(q, e, q_2)$".

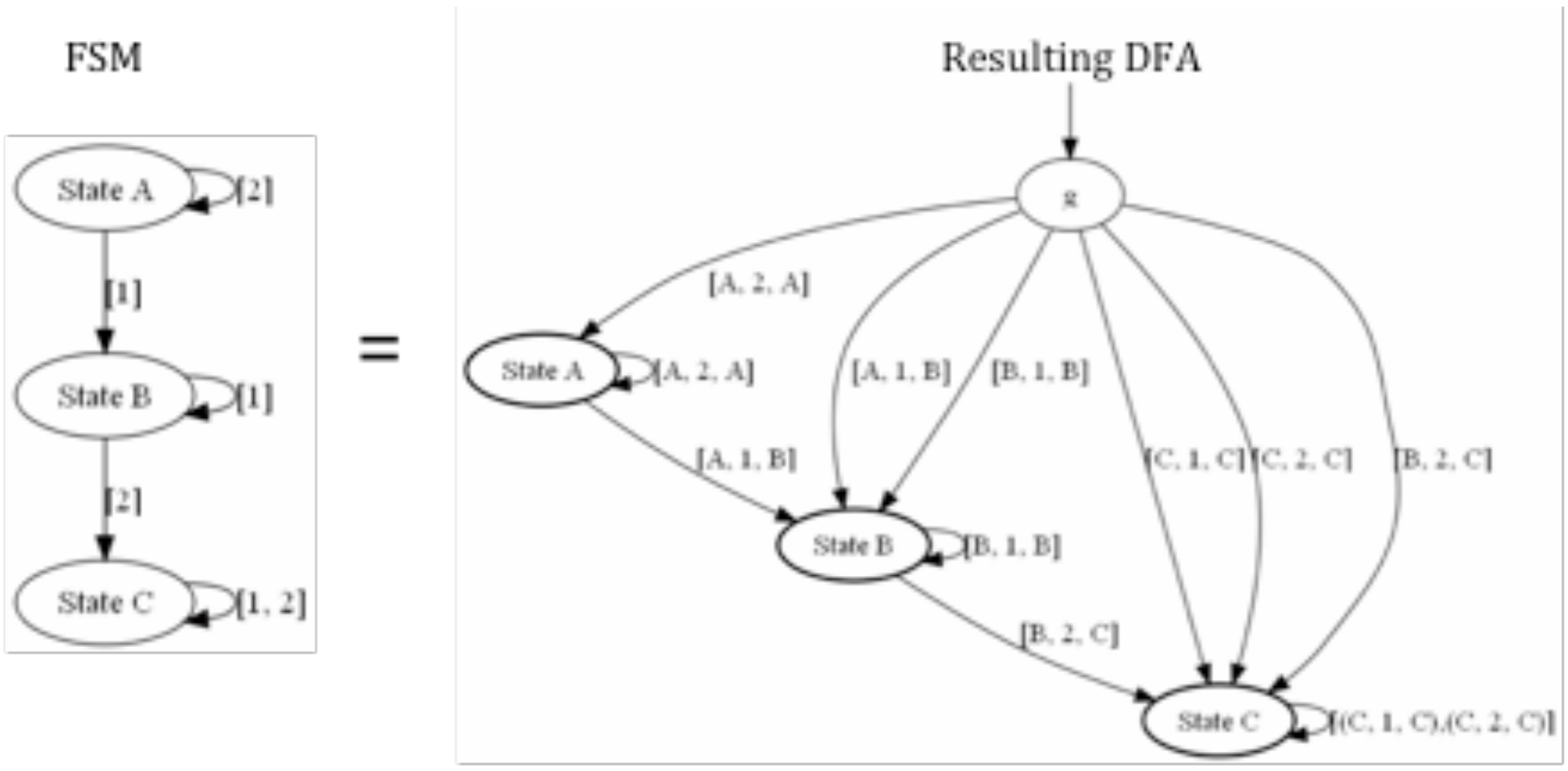

**Figure 2 – DFA that represents the computations of the given FSM**

***Define:*** $\delta_1 : Q_1 \times \Sigma_1 \rightarrow Q_1$
such that:
$\forall e \in \Sigma, \forall q, q_2 \in Q : ((q, e), q_2) \in \delta \Leftrightarrow (q, (q, e, q_2), q_2) \in \delta_1 \wedge (g, (q, e, q_2), q_2) \in \delta_1$

4. Define the set of accepting states $F_1$ as the set of states $Q$.
***Define:*** $F_1 = Q$

The definition of acceptance in the model is any sequence of transitions for which $\delta_1$ returns a value.

When done, if there existed a state machine with three states (A, B, C) whose possible input symbols were (1, 2), a model could be created that accepts its computations as strings (figure 2). The possible events being $\Sigma_1$ = {[A,1,B],[A,2,A],[B,2,C],[B,1,B],[C,1,C],[C,2,C]}. In this model the process of moving from state A to state C could be represented as a string: [A,1,B][B,2,C]. If the start state is unknown, which is common in real world systems, the model also accepts an existing transition from start state *g*. This allows a string such as [B,1,B][B,2,C] to be accepted even if the start state was not explicitly defined as state B.

The resulting automaton ($M_1$) represents the set of computations of *M*. Valid computations of M, are defined as:

$$C_M \subseteq \Sigma_1^*, \text{ where } s \in C_M \Leftrightarrow (\forall_i\ |s|-1 \geq i \geq 1 : \psi_q(s_{i+1}) = \psi_{q2}(s_i))$$

**Observations**
Observations are fragments of system computations observed by a witness. While it is possible to observe a full transition sequence, a witness may also only have a partial knowledge of such a transition. For this reason an unknown or unobserved element in an observation triple is denoted by (*?*). For example, if there existed a transition [A,1,A] and [A,1,B] from state A, but the witness did not observe the resulting state of the transition, the observation would be given as [A,1,?].

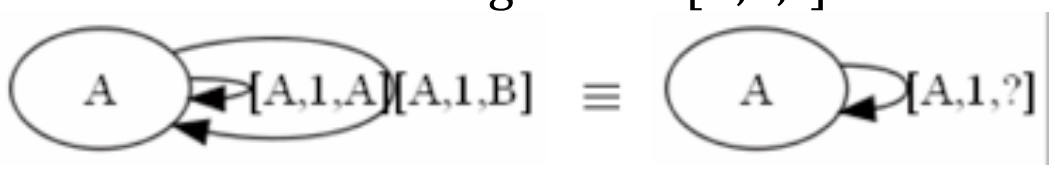

Figure 3 - An Observation with an Unknown Element

Likewise, if no transition or sequence of transitions were observed for a point in time, it is defined as no-observation and is represented by the triple [?,?,?]. A no-observation means that any transition could have occurred from any possible state to any other state on any possible input. A no-observation provides no restriction to the possible pattern of triples.

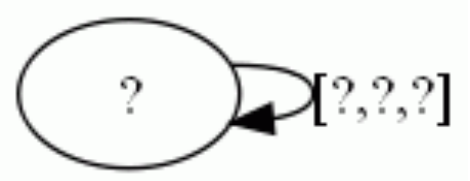

Figure 4 - Transition from any state to any other state on any transition

**Witness Statements**
Witness statements may be represented as patterns over the sequences of transition triples. These patterns are defined by ordering a group of observations chronologically. This creates partial views of system computations observed by a witness. For example, if a witness observed a system in state A then after some time observed the same system in state B, the observations could be described as [A, ?, ?] and [?, ?, B]. Ordering these observations chronologically and filling in gaps with no-observations, creates a sequence of transitions triples that represents the observed states, and each possible transition between these states.

[A, ?, ?][?, ?, ?][?, ?, B]

It is possible to create a finite state machine that accepts exactly the constructed string.

**III. Analysis of the Evidence**
This section illustrates how to derive the feasibility of a given witness statement by using the proposed method. The printer analysis case explored in (Gladyshev, Formalising Event Reconstruction in Digital Investigations, 2004)(Gladyshev & Patel, Finite State Machine Approach to Digital Event Reconstruction, 2004) will be examined.

**The Case** (adopted from (Gladyshev & Patel, Finite State Machine Approach to Digital Event Reconstruction, 2004))
The local area network at ACME Manufacturing consists of two personal computers and a networked printer. Its two users, Alice (A) and Bob (B), share the cost of running the network. Alice, however, claims that she never uses the printer and should not be paying for the printer consumables. Bob disagrees; he says that he saw Alice collecting printouts. The system administrator, Carl, has been assigned to investigate this dispute.

To get more information about how the printer works, Carl contacted the manufacturer. According to the manufacturer, the printer works as follows:

1. When a print job is received from the user it is stored in the first unallocated directory entry of the print job directory.
2. The printing mechanism scans the print job directory from the beginning and picks the first active job.

3. After the job is printed, the corresponding directory entry is marked as “deleted”, but the name of the job owner is preserved.

The manufacturer also noted that

4. The printer can accept only one print job from each user at a time.
5. Initially, all directory entries are empty.

After that, Carl examined the print job directory. It contained traces of two of Bob's print jobs, and the rest of the directory was empty:

job from B (deleted)
job from B (deleted)
empty
empty
empty
...

**Informal Analysis**

Carl reasons as follows: If Alice never printed anything, only one directory entry must have been used because the printer accepts only one print job from each user. However, two directory entries have been used and there are no other users except Alice and Bob. Therefore, it must be the case that both Alice and Bob submitted their print jobs at the same time. The trace of Alice's print job was overwritten by Bob's subsequent print jobs. Given the printer model, it would be possible to test Carl’s reasoning process, and explore other possibilities Carl’s informal reasoning may have missed.

**Defining the Printer Model**

Carl’s observation claims that there were traces of print jobs in only the first two printer queues. Since the remaining queues were observed as being empty, there is no other relevant information that could be derived from them. Because of this, the investigation and modeling can be defined in terms of only the first two queues. This helps to reduce the scope of possibilities, and reduces the complexity required to model the system.

Each queue can be defined as possibly having the following states:

- E – the queue is completely empty
- A – the queue contains a job from Alice
- B – the queue contains a job from Bob
- Del_A – the queue contains a deleted job from Alice
- Del_B – the queue contains a delete job from Bob

The possible transitions that can act on the queues are:

- Add_A – represents Alice printing
- Add_B – represents Bob printing
- Take – represents the deleting the job

The printer, as defined in the case, will be represented as $Printer = (Q, \Sigma, \delta)$ (Gladyshev & Patel, Finite State Machine Approach to Digital Event Reconstruction, 2004), where

- $Q$ = {{E, A, B, Del_A, Del_B} × {E, A, B, Del_A, Del_B}}
- $\Sigma$ = {Add_A, Add_B, Take}
- $\delta : Q \times \Sigma \rightarrow Q$ is a transition function that determines the next state

By applying the proposed algorithm, the computations of *Printer* can be modeled. This is represented as $Printer_1 = (Q_1, \Sigma_1, \delta_1, g, F)$, where

- $Q_1$ = {({E, A, B, Del_A, Del_B} × {E, A, B, Del_A, Del_B}), $g$}
- $\Sigma_1 = \{\varphi \mid \varphi = (q, e, q_2)\}$
- $\delta_1 : Q_1 \times \Sigma_1 \rightarrow Q_1$ is a transition function that that determines the next state of $M_1$
- $g$ is the start state where $g \notin Q$
- $F$ is the set of accepting states equal to the set $Q$

The resulting model of $Printer_1$ is represented in figure 5.

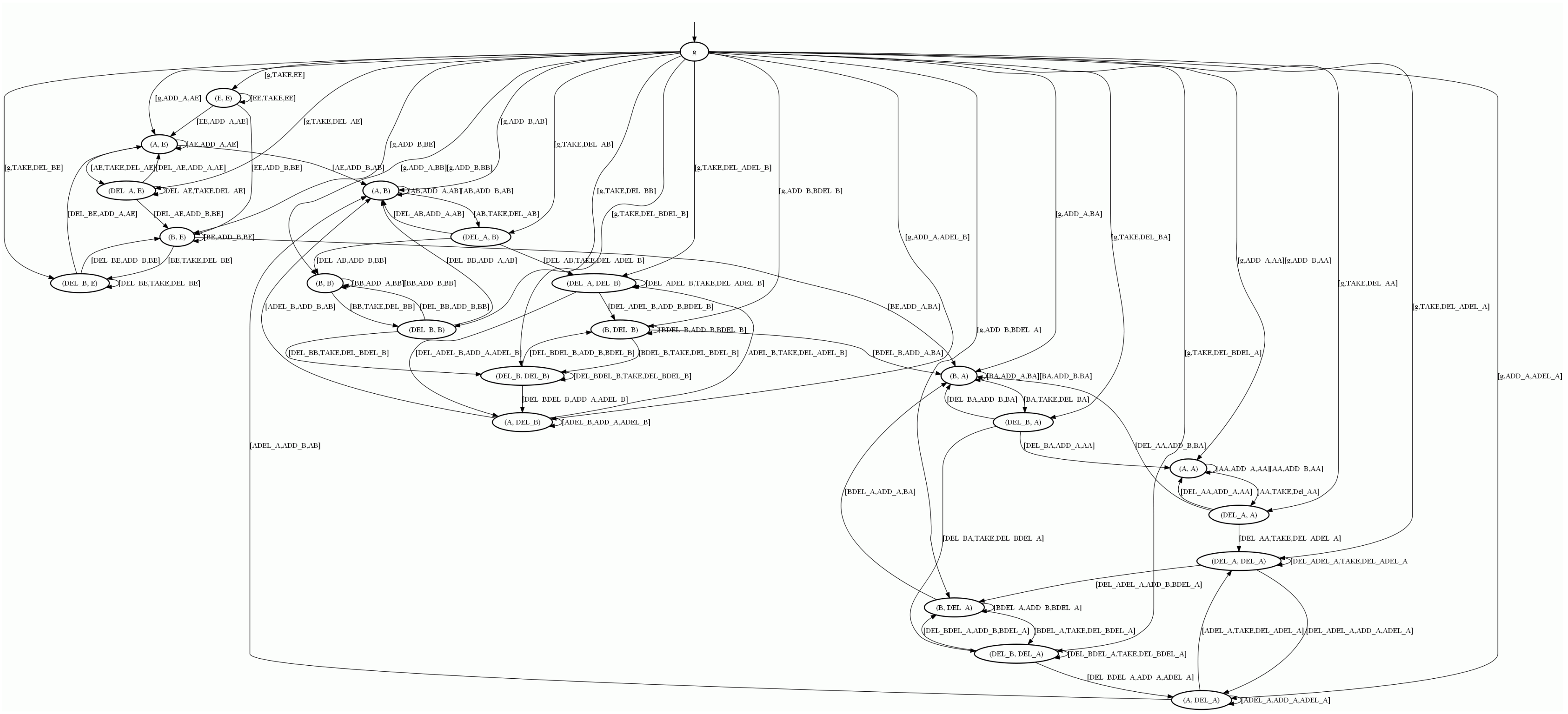

**Figure 5 - Model of *Printer₁* with generic start state *g***

**Restriction of the Model**
Given this definition of $Printer_1$, all possible combination of states and transitions are represented for the system. Applying known, definite observations can restrict the number of possible events. If the observations come from a trusted source, such as a direct observation by the investigator, then the resulting restricted model can also be trusted to be accurately representative of the actions of the system.

The first known observation is that of the manufacturer who claims the initial state of the printer is (E, E), meaning that both queues were empty. The manufacturer has no knowledge after shipping. This observation can be defined as:

[(E,E), *?, ?*]

The second observation is that of Carl, who observed the final state as (Del_B, Del_B), meaning that there are two deleted jobs from Bob in the queue. He has no knowledge of the queues before the investigation.

[*?, ?,* (Del_B, Del_B)]

These observations can be combined to create a restricting witness statement. Unknown elements within the observation are denoted by (?)

$ws_1$ = [(E,E), *?, ?*][*?, ?, ?*][*?, ?,* (Del_B, Del_B)]

$ws_1$ says that the printer queues start in state (E,E), transitioned through any state any number of times, and ended in state (Del_B, Del_B). This statement can be intersected with $Printer_1$ to produce a model that *both* state machines accept. The restricted model is shown as $Printer_{Restricted}$.

The actual restriction in this instance comes from setting the start state. This not only removes the generic start state *g*, but also restricts states other than (E,E) from being the first in the sequence. This statement also sets the only accepting state to (Del_B, Del_B) meaning that only stings ending in this state will be accepted. As shown in Figure 6, $Printer_{Restricted}$ contains fewer possible transitions than the full, unrestricted printer model in figure 5.

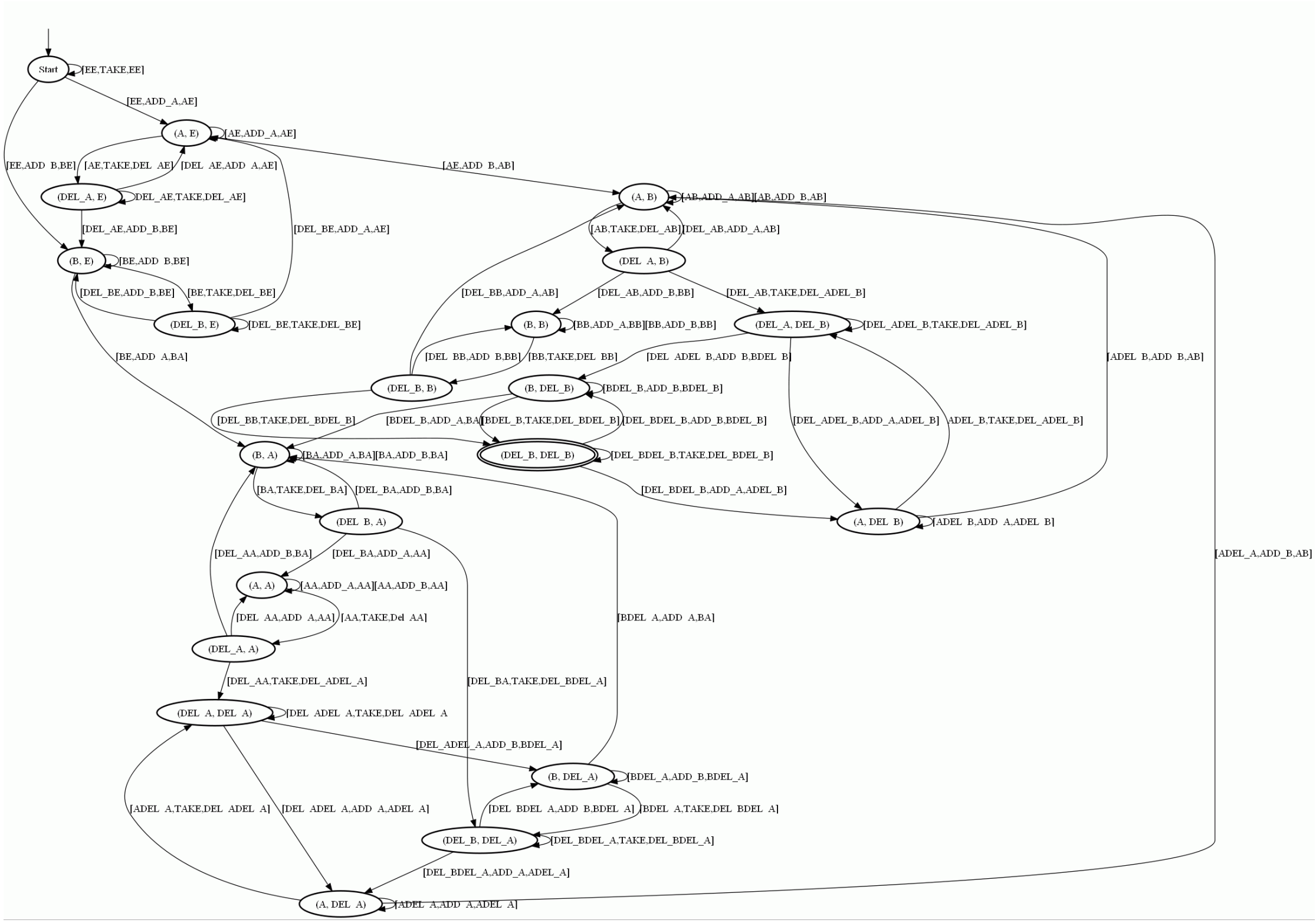

**Figure 6 - *$Printer_1$* intersected with *$ws_1$***

**Modeling uncertain Witness Statements**
Next the witness statement for Bob is

created. His statement is simply that both he and Alice used the printer. Essentially this puts no restriction on the constructed system. In terms of transitions, Bob claims that any transition could have happened any number of times.

$$ws_{Bob} =$$
$$([?, \text{Add_A}, ?] \mid [?, \text{Add_B}, ?] \mid [?, \text{Take}, ?])$$

This claim could easily be expressed as the regular expression

$$([?, \text{Add_A}, ?]|[?, \text{Add_B}, ?]|[?, \text{Take}, ?])^*$$

The resulting automaton is a single state that accepts the inputs ‘Add_A’, ‘Add_B’, and ‘Take’ from any state.

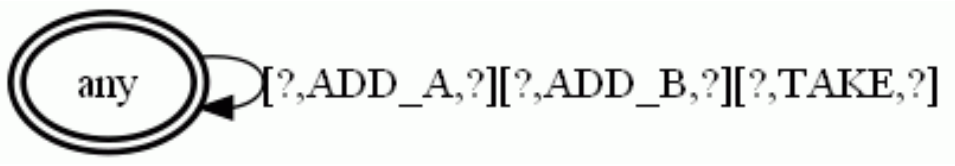

Figure 7 - Model of Bob's Statement

Next the witness statement for Alice is created. Her statement is simply that she did not print. This can be explained as any transition except ‘Add_A’, and can be expressed as:

$$ws_{Alice} = ([?, \text{Add_B}, ?] \mid [?, \text{Take}, ?])$$

This claim could also be easily expressed as the regular expression

$$([?, \text{Add_B}, ?] \mid [?, \text{Take}, ?])^*$$

The resulting automaton is a single state that accepts inputs ‘Add_B’ and ‘Take’, but not ‘Add_A’.

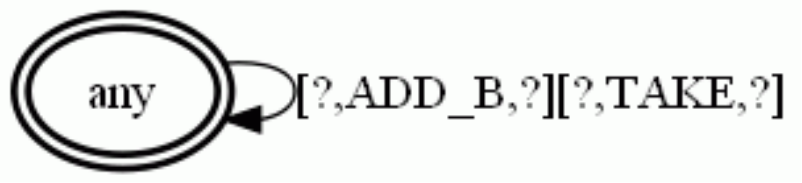

Figure 8 - Model of Alice's Statement

Because every transition in $Printer_{Restricted}$ is also present in $ws_{Bob}$, intersecting will cause no reduction in states. This means that the final observed state (Del_B, Del_B) is *possible* with Bob’s statement.

Alice’s statement, however, does restrict the possible transitions of $Printer_{Restricted}$ (Figure 9). Because of this restriction the final state (Del_B, Del_B) is not reachable. This means that the statement given by Alice is not possible in accordance with the system model, meaning that she must be lying.

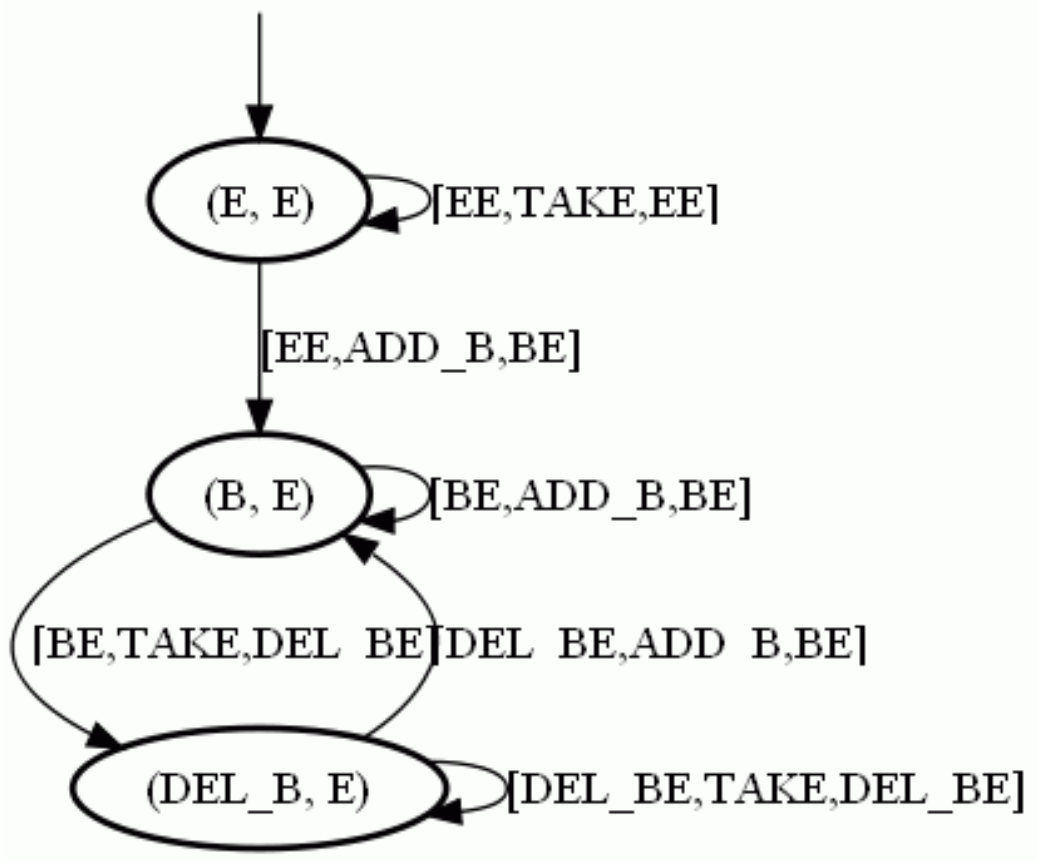

Figure 9 - Result of Intersection between the Printer Model and Alice's Statement

## IV. Conclusions

In the given printer case it is shown that by modeling the system and witness statements it is possible to computationally find whether the given statement could have happened. The printer case worked well due to the fact that the witness statements could be defined as restrictions on system functions. It is possible, even given specific observations, that the witness statement model may not restrict transitions of the system. For example, a witness statement where the first observed state was A, followed by unknown observations, and a last observed state C. These observations can be described as:

[A, ?, ?] – first observed
[?, ?, C] – last observed
Gaps in the story are represented by no-observations, [?, ?, ?].

The new witness statement is

$$ws = [A, ?, ?][?, ?, ?][?, ?, C]$$

which is illustrated in figure 10. The issue with this type of statement is the fact that between states A and C, any event could have happened any number of times. If the statement is too general then there will be no state reduction.

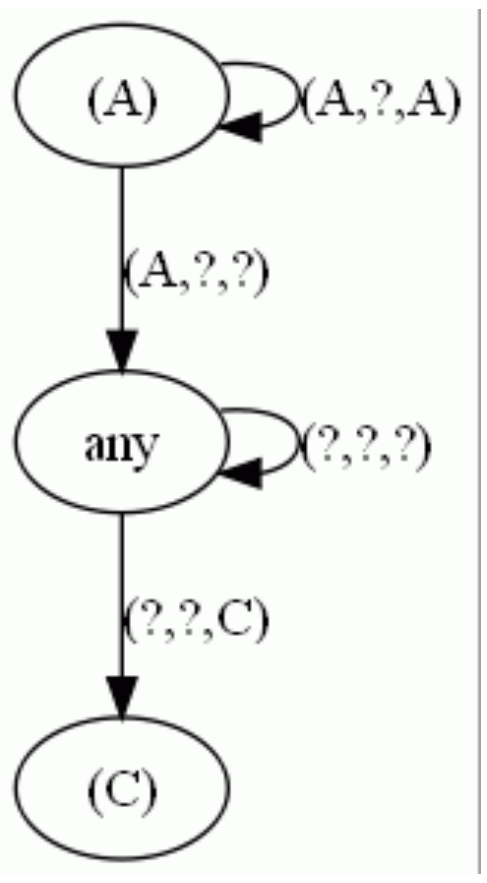

Figure 10 – General Witness Statement

For example, figure 11 shows a slightly more complex system model where each state is reachable from every other state. In this case, intersecting the witness statement (figure 10) with this model does not provide any restriction if an unknown input (*?*, *?*, *?*) is accepted. This is because each possible transition in the system model is reachable between states A and C.

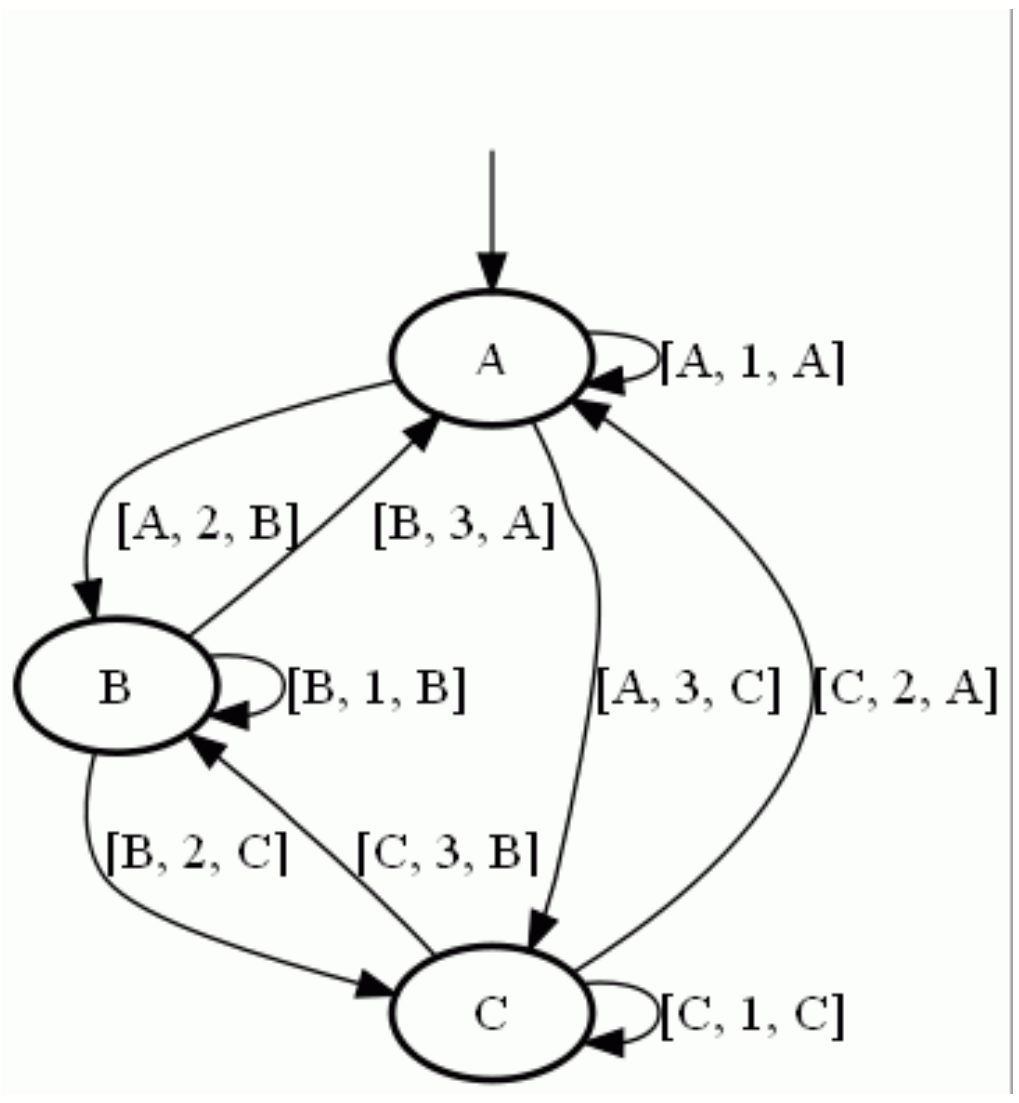

Figure 11 – Simple System Model

The advantage of state machine analysis is the fact that each possibility *is* considered. By modeling statements given by a witness as state machines, it then becomes possible to test whether there is a string (piece of evidence) that is accepted by both FSMs. This allows a given witness statement to be tested against the system itself to determine if it is computationally possible. However, the downfall of FSM analysis continues to be the fact that the analysis of real systems remains impractically susceptible to exceedingly large state spaces. This makes the modeling of even the simplest real systems challenging. As such, any practical application would have to focus on an extremely specific sub-system.

This method is also limited in the fact that, while it can detect if something is impossible, it cannot detect if something definitely happened. This leads to issues if the suspect's statement is exceedingly general. In this case much more evidence would be needed to reduce the possible states of the system model. This also means that the approach is only suitable to help the defense cast doubt by finding alternative possibilities that agree with the evidence.

**Applications and Future Work**

For the practical application of this method the investigator must focus on simple systems or be able to greatly generalize the system in question. The modeling of some components of a computer system could be automated, however since computers are a collection of sub-systems interacting with each other, some level of abstraction would be needed to derive only the operations unique to a given sub-system. There has yet to be a rigorous method for determining the appropriate level at which to model complex systems as FSMs that would be general enough for practical event reconstruction purposes. This is the focus of future research. Given the complexity of real systems, and even sub-systems, one possible way to proceed appears to be modeling a collection of sub-systems and grouping them to represent processes of the overall system. It is not clear, however, how to reconstruct these systems in a post-modem analysis. Another approach involves simplification of the system by focusing on events that are known to have definitely happened rather than reconstructing all possible events. These could be user or system actions that can be proved to have definitely happened due to traces in the system. While these events alone may not necessarily be substantial proof of an event, a collection of definitely happened events could be modeled and combined with the proposed method to produce a minimal system of relevant events that witness statements could be tested against.